\documentclass[11pt,a4paper]{article}
\usepackage[affil-it]{authblk}

\usepackage[T1]{fontenc}
\usepackage{lmodern}

\usepackage{graphicx}
\usepackage{booktabs}
\usepackage{amsmath}
\usepackage{hyperref}

\usepackage{listings,xcolor}

\usepackage{xurl}
\newcommand{\codepath}[1]{\nolinkurl{#1}}

\lstdefinestyle{yaml}{
  basicstyle=\ttfamily\footnotesize,
  rulecolor=\color{black},
  string=[s]{'}{'},
  stringstyle=\color{blue},
  comment=[l]{:},
  commentstyle=\color{black},
  morecomment=[l]{-},
  showstringspaces=false,
  keepspaces=true,
  columns=fullflexible
}

\title{FAST-HEP: Compiling Declarative Analysis Workflows for
High-Energy Physics and Beyond}
\author{Luke Kreczko}
\affil{University of Bristol, UK}
\date{August 2026}

\begin{document}

\maketitle

\begin{abstract}
  High-energy physics analyses increasingly rely on complex software
  workflows whose scientific lifetime often exceeds that of the
  underlying software ecosystem. Maintaining reproducibility while
  accommodating evolving analysis software, data formats, and
  execution environments therefore remains a significant challenge.
  These challenges are not unique to high-energy physics and are
  shared by many data-intensive scientific analyses.

  We present FAST-HEP and its workflow engine, Flow, which combines a
  declarative workflow language, compiler, and runtime. Flow
  separates the scientific description of a workflow from its
  implementation and execution, and compiles workflows into
  backend-independent execution plans through normalization, graph
  construction, dependency analysis, and execution planning. A common
  runtime then orchestrates the resulting plan using replaceable
  capabilities. This architecture enables static validation and
  modular replacement of analysis operations, execution backends, and
  storage technologies, while recording provenance throughout
  compilation and execution.

  Although developed for the requirements of high-energy physics
  analysis, Flow's workflow model and orchestration layer are
  domain-independent. By applying compiler techniques to scientific
  analysis workflows, FAST-HEP provides a foundation for workflows
  that are transparent, extensible, portable, and reproducible,
  allowing scientific analyses and their supporting software
  ecosystems to evolve independently.
\end{abstract}

\section{Introduction}
\label{sec:introduction}

High-energy physics analyses are long-lived scientific products built on a
rapidly changing software and computing ecosystem. During the lifetime of an
analysis, data-processing libraries, storage technologies, computing
infrastructure, and the researchers maintaining it may all change. Increasing
data volumes and increasingly heterogeneous computing resources create further
pressure for the software to evolve. Scientific reproducibility therefore
requires more than preserving analysis code: an analysis must remain
understandable, maintainable, and executable as its surrounding technology and
personnel change. We refer to this broader requirement as \emph{analysis
sustainability}: preserving both the software and a sufficiently explicit
description of the scientific workflow for its results to be understood,
reproduced, and evolved.

In practice, the complete description of a HEP analysis workflow is
rarely captured by a single program or configuration file. Scientific
choices may be distributed
across analysis code, dataset definitions, corrections and calibrations,
systematic variations, job-submission scripts, software environments, and
documentation. Some of the knowledge connecting these pieces may remain
implicit within the research team. This is particularly challenging when much
day-to-day analysis development is performed by PhD students and other
early-career researchers, whose involvement may be shorter than the lifetime
of the analysis. When researchers move on, it can become difficult to establish
not only how to rerun an analysis, but precisely which datasets, corrections,
systematic variations, and software configurations contributed to a particular
result.

The HEP community has developed important infrastructure addressing different
parts of this reproducibility problem. REANA~\cite{Simko2019REANA} provides
mechanisms to preserve and re-execute computational workflows, while
HEPData~\cite{Maguire2017HEPData} provides long-term publication and
preservation of data and auxiliary material associated with scientific
results. Such efforts make preservation considerably more systematic, but
curation remains an additional task and depends on the information available
to preserve. Analysis sustainability therefore benefits from making scientific
intent, dependencies, and workflow information explicit during normal analysis
development rather than reconstructing them only at publication time.

Even when this information is captured in software, scientific intent can be
difficult to separate from its implementation. HEP analyses have traditionally
been expressed imperatively, a model that maps naturally onto event-based data
but can couple the analysis definition to control flow, library interfaces,
data representations, and execution assumptions. A conceptually simple
scientific statement may consequently be spread across substantial amounts of
implementation and supporting code, making both understanding the analysis
and changing its underlying technology more difficult.

FAST-HEP, originally the Faster Analysis Software Taskforce, emerged in 2017
from a UK-led collaboration exploring how developments in the scientific Python
ecosystem could improve HEP analysis. Interfaces such as
\codepath{root_numpy}~\cite{Dawe2017RootNumpy} and
\codepath{root_pandas}~\cite{chris_burr_2019_3348073} were attracting interest
as ways of exposing data stored in ROOT~\cite{Brun1997ROOT} to the
array-oriented Python ecosystem. They opened an attractive route towards new
analysis tools, but adopting them involved more than learning a new API:
analysts accustomed to event-by-event loops had to reformulate computations as
operations on arrays and columns. FAST-HEP used short, focused
``hack-shops'' to explore such technologies while asking a broader question:
could analysts describe the scientific workflow they wanted without also
having to adopt the programming and execution model of each underlying
technology?

FAST-HEP therefore explored a declarative approach to workflow description
that shifted the author's focus from \emph{how} a computation should be
implemented towards \emph{what} should be computed. Selections,
derived quantities, histograms, and other
analysis operations described the intended computation while the underlying
software determined how to realise it. Besides lowering the barrier to adopting
new tools, this separation offered opportunities for optimisation, portability,
and reproducibility: in principle, the same scientific description could be
interpreted using different libraries or computing infrastructures.

Over the following years this model developed into the FAST-HEP tool\-set and
was applied to real analysis workflows. By 2019, FAST-HEP was being used in CMS
and the LUX-ZEPLIN experiment, as well as by students working with other
experiments~\cite{Krikler2020FASTHEP}. These applications demonstrated that a
concise declarative description could be useful across different analyses,
data structures, and experiments. Development, however, remained largely
community-driven, and an important limitation gradually became apparent:
although the author-facing description separated scientific intent from
implementation, the software components underneath it remained more tightly
coupled than the interface suggested.

In 2021, FAST-HEP received an IRIS Digital Research Infrastructure (DRI)
funding award (STFC grant ST/W00495X/1) providing six months of dedicated
Research Software Engineer effort to modernise the software stack. This
provided a practical test of a central assumption of the project: that
underlying technologies could evolve while the declarative analysis description
remained stable. The planned work included migration to new generations of
Uproot~\cite{Pivarski2017Uproot} and
Awkward Array~\cite{Goyal2023Awkward},
replacement of the histogramming infrastructure, support for more complex and
structured data, and, in particular, enabling the use of GPUs. In
practice, the modernisation required substantially more
development effort than anticipated. Components that appeared
conceptually independent had accumulated dependencies on shared data
representations, library interfaces, and execution assumptions. As a
result, apparently local changes propagated across the software stack.

The modernisation exposed an important distinction between a declarative
interface and a modular architecture. Separating the analysis description from
imperative code had not made the implementations underneath it independently
replaceable. This had a human as well as a technical cost: for a project
sustained largely through volunteer contributions and short periods of focused
development, experiments that required framework-wide knowledge and changes
were difficult to sustain and discouraged contributor engagement. Independent
replaceability and low-cost experimentation therefore became requirements for
software sustainability rather than merely desirable implementation properties.

These lessons motivated the development of Flow, the workflow system
presented in this paper.
Scientific workflows are described independently of
the concrete implementations used to realise them. Compilation
transforms the workflow description into a normalised workflow
and logical graph, analyses and validates its dependencies, and produces an
inspectable execution plan.
Implementations, instrumentation, and
execution environments can then evolve independently of the scientific
workflow description. The intended boundary is therefore no longer simply
between declarative configuration and imperative code, but between a stable
description of scientific intent and an evolving ecosystem of software and
computing technologies.

This separation has become more important as that ecosystem has continued to
diversify. Analyses increasingly span local, distributed, and heterogeneous
computing resources while their underlying software continues to evolve.
AI-assisted software development adds another dimension: producing software
is becoming easier without necessarily making the resulting workflow easier
to understand or verify. This makes explicit, inspectable, and reproducible
scientific workflows more important, not less. Flow therefore aims to
be concise for authors while
compilation provides the explicit structure required for validation,
provenance, optimisation, and execution.

The remainder of this paper describes how these principles are realised in
Flow. Section~\ref{sec:architecture} introduces the architecture and extension
model. Section~\ref{sec:workflow-language} presents the author-facing workflow
language, followed by compilation into a backend-independent execution plan in
Section~\ref{sec:compilation}. Section~\ref{sec:execution} describes how the
plan is mapped onto execution environments. Finally,
Section~\ref{sec:lessons} discusses the experience and lessons that
have shaped the design, before we conclude in
Section~\ref{sec:conclusions}.

\section{Architecture}
\label{sec:architecture}

Flow is designed around the principle that scientific workflows evolve on a
different timescale from the software used to execute them. New libraries,
data representations, execution environments, and computing technologies
should therefore be introducible without requiring authors to reformulate
their scientific workflow. Replaceability is consequently treated as an
architectural requirement rather than an implementation detail.

The architecture achieves this through explicit boundaries separating workflow
authoring, compilation, execution, and computing infrastructure. Rather than
interacting directly with one another, these components communicate through
well-defined representations and contracts. Each boundary isolates concerns
that evolve independently while exposing the information required by the next
stage of the workflow lifecycle.

These boundaries deliberately extend beyond Flow itself. Most domain-specific
functionality, including analysis operations, data sources, compiler
extensions, and execution backends, is intended to be supplied by external
packages. Flow provides the framework that composes these capabilities rather
than implementing them directly.

\begin{figure}[!htbp]
  \centering
  \includegraphics[width=0.8\linewidth]{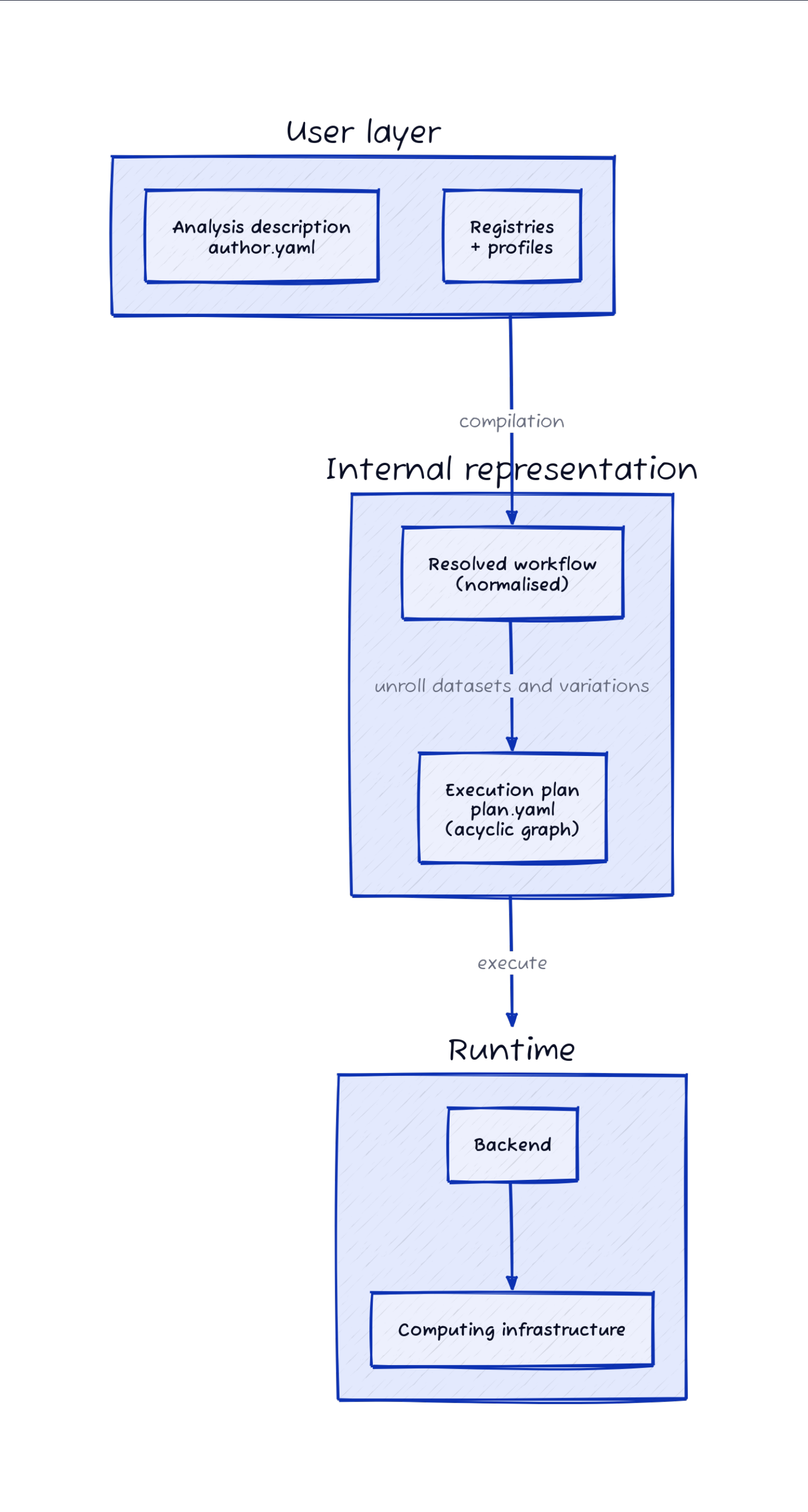}
  \caption{
    Principal architectural boundaries in Flow. Compilation progressively
    transforms an author-facing workflow into increasingly explicit
    representations before execution. Capabilities used during compilation and
    execution can be supplied or replaced independently.
  }
  \label{fig:architecture}
\end{figure}

Figure~\ref{fig:architecture} illustrates the overall architecture. The
central path represents a progression of increasingly explicit workflow
representations rather than a fixed stack of implementations. Compilation
transforms an author-facing workflow description into an executable plan,
which is subsequently interpreted by the runtime on the selected execution
backend.

\subsection{From workflow to execution}
\label{sec:architecture-representations}

Flow treats scientific workflows as programs that are compiled rather than
interpreted directly during execution. Compilation progressively transforms a
concise author-facing workflow into increasingly explicit representations,
culminating in a backend-independent execution plan.

Conceptually, this progression is

\begin{center}
  \begin{tabular}{c}
    Workflow description \\
    $\downarrow$ \\
    Normalised workflow \\
    $\downarrow$ \\
    Logical graph \\
    $\downarrow$ \\
    Execution plan \\
    $\downarrow$ \\
    Runtime
  \end{tabular}
\end{center}

Each representation captures a different stage of understanding the workflow.
The workflow description expresses scientific intent in a concise form suitable
for authors. Compilation progressively makes inherited configuration,
dependencies, and execution structure explicit, producing representations that
are increasingly suitable for analysis, validation, and execution.

These intermediate representations are intentionally inspectable and
serialisable. They provide stable architectural boundaries between compiler
phases while also serving as useful artifacts for debugging, comparison,
validation, and provenance.

\subsection{Compilation and execution}
\label{sec:architecture-compilation-runtime}

Compilation and execution have distinct responsibilities.

The compiler transforms the workflow description into an execution plan. During
this process it resolves the workflow environment, analyses dependencies,
validates the workflow, and constructs the information required for execution.
These analyses are performed before any scientific computation begins.

The execution plan forms the principal architectural contract between the
compiler and the runtime. Once a compatible plan has been produced, the runtime
does not need to know how the workflow was authored or how it was compiled. It
receives an explicit description of the work to perform together with the
dependencies between its constituent operations.

The runtime is responsible for interpreting this plan, executing its
operations, managing execution lifecycles, and recording provenance. Execution
backends are in turn responsible for mapping the plan onto a particular
computing environment, whether local or distributed. Scheduling, resource
allocation, and worker management therefore remain separate from the scientific
workflow itself.

\subsection{Extension boundaries}
\label{sec:architecture-extensions}

The architectural boundaries also define the primary extension points within
Flow. Rather than treating all extensions as interchangeable plugins, Flow
classifies extensions according to the architectural role they perform.

Workflow extensions provide the functionality available to scientific
workflows. Compiler extensions contribute additional analyses or artifacts
during compilation. Runtime extensions determine how compiled workflows are
executed or mapped onto particular computing environments.

This classification allows each part of the system to evolve independently
while making the purpose of every extension explicit. New functionality can be
introduced at the architectural boundary corresponding to its responsibility
without modifying unrelated parts of the framework.

The following chapters describe the concrete extension mechanisms associated
with each architectural layer.

\subsection{Operations, registries, and profiles}
\label{sec:architecture-operations}

Scientific workflows refer to named operations rather than directly to
libraries, classes, or functions. Each operation is defined through an
operation specification that describes the information visible to the compiler,
while one or more implementations provide the executable behaviour used by the
runtime.

Registries provide the indirection between workflow descriptions and available
implementations. Packages register operations together with their
specifications, allowing Flow itself to remain independent of domain-specific
software.

Profiles assemble these registrations into a particular analysis environment.
A profile can combine functionality from several packages, provide
configuration, and select alternative implementations without modifying the
scientific workflow itself. The same workflow can therefore be compiled in
different software environments while preserving its scientific intent.

This separation allows scientific workflows to depend on stable architectural
contracts rather than concrete implementations. Implementations can evolve,
be replaced, or be compared experimentally without requiring changes to the
workflow description.

\section{Workflow language}
\label{sec:workflow-language}

The Flow authoring language provides a declarative description of a scientific
workflow rather than an execution script. Authors describe the data entering
the analysis, the operations to apply, relationships between analysis stages,
and the outputs they wish to produce. Details such as dependency resolution,
partitioning, scheduling and implementation selection are resolved during
compilation.

A workflow can therefore be read primarily as a description of the scientific
analysis itself. The following sections introduce the language using a complete
running example.

\subsection{A complete workflow}
\label{sec:workflow-language-example}

Listing~\ref{lst:nasa-workflow} shows a complete workflow using the current
authoring syntax. The example reads a NASA exoplanet dataset from Parquet,
expands the nested planet information into rows, selects planets with radii
between 0.8 and 1.2 Earth radii, projects a small set of fields, and writes a
summary table.

The syntax shown here should be understood as a snapshot of the current Flow
interface rather than a fixed language specification. In particular, further
compiler development is expected to reduce information that authors need to
state explicitly. The underlying model - datasets, data sources, analysis
operations, outputs, and their relationships - is intended to remain stable
even as the author-facing representation becomes more concise.

\begin{lstlisting}[
  caption={A complete Flow workflow using the NASA exoplanet example.},
  label={lst:nasa-workflow},
  style={yaml},
]
version: 1.0

use:
  profiles:
    - registry
    - fasthep_workshop:registry

data:
  datasets:
    - name: nasa_exoplanets
      files: [data/NASA/exoplanets.parquet]
      eventtype: data
  defaults:
    eventtype: data
    tree_primary: planets

sources:
  planets:
    kind: workshop.parquet
    stream_type: event_stream

analysis:
  stages:
    - id: PlanetRows
      op: workshop.tabular.explode
      params:
        fields:
          - planet_name
          - planet_radius
          - planet_period
        keep_fields:
          - name

    - id: EarthSizedPlanets
      op: workshop.tabular.filter
      params:
        expr: "(planet_radius > 0.8) & (planet_radius < 1.2)"

    - id: PlanetTable
      op: workshop.tabular.project
      params:
        fields:
          - name
          - planet_name
          - planet_radius
          - planet_period
      write:
        - kind: workshop.console_table
          path: snippets/planets.txt
          when: final
          fields:
            - name
            - planet_name
            - planet_radius
            - planet_period
          columns:
            - template: "{name} {planet_name}"
              header: Planet
            - field: planet_radius
              header: Radius [Earth radii]
              format: ".3f"
            - field: planet_period
              header: Period [days]
              format: ".3f"
          sort_by:
            - planet_radius
            - name
            - planet_name
          limit: 12
\end{lstlisting}

The top-level structure separates concerns that are often interleaved in
analysis code. The \codepath{use} block selects the profiles that provide the
available functionality; \codepath{data} describes the datasets to analyse;
\codepath{sources} describes how those datasets enter the data-flow graph; and
\codepath{analysis} describes the operations applied to the resulting products.

Although Listing~\ref{lst:nasa-workflow} appears relatively long, the
computational path
itself consists of only three stages: exploding the nested planet records,
filtering by radius, and projecting the fields required for the final table.
Much of the remaining configuration describes the dataset, software
environment and output generation. These concerns remain explicit without
being embedded in the implementation of the analysis operations.

\subsection{Data and analysis stages}
\label{sec:workflow-language-stages}

The workflow separates the description of input data from the mechanism used
to introduce those data into the analysis. In
Listing~\ref{lst:nasa-workflow}, the \codepath{data} block identifies the
\codepath{nasa\_exoplanets} dataset and its files, while the \codepath{sources}
block selects the functionality that turns those files into an
\codepath{event\_stream}. A dataset is therefore not tied to a particular file
reader or data representation.

Once data have entered the workflow, the \codepath{analysis} block describes
successive operations on the resulting products. The NASA example contains
three stages: \codepath{PlanetRows} expands the nested planet information,
\codepath{EarthSizedPlanets} applies the radius selection, and
\codepath{PlanetTable} retains the fields needed for the final output. Each
stage names an operation through \codepath{op} and supplies its configuration
through \codepath{params}.

The structure of a stage is part of the Flow language, but the meaning of its
parameters belongs to the registered operation. For example, Flow itself does
not need to define the syntax of the expression used by
\codepath{workshop.tabular.filter}, or the meaning of the field lists accepted
by \codepath{workshop.tabular.project}. The corresponding operation
specification exposes the information required during compilation, including
which configured values represent dependencies or newly produced fields.

This provides an important boundary between the workflow language and
domain-specific functionality. New operations can introduce configuration
appropriate to their domain without requiring new syntax in Flow itself, while
their specifications still allow the compiler to reason about how those
operations compose.

\subsection{Steering the workflow}
\label{sec:workflow-language-steering}

Not every operation in an analysis necessarily applies to every input dataset.
Flow therefore allows applicability to be expressed as part of the workflow
rather than hidden inside the implementation of an operation. For example, a
stage using generator-level information might be restricted to simulated data:

\begin{lstlisting}[style=yaml]
- id: GeneratorWeights
  op: hep.generator_weights
  applies_to:
    eventtype: mc
  params:
    ...
\end{lstlisting}

During compilation, such conditions are evaluated in the context of each
dataset. Operations that do not apply are omitted from the corresponding
execution path. A workflow can therefore describe data- and
simulation-specific processing explicitly without requiring operations to
contain their own hidden checks or authors to maintain separate analysis
definitions.

\subsection{Variations}
\label{sec:workflow-language-variations}

Many scientific analyses require several closely related versions of the same
computation. In HEP these commonly correspond to systematic uncertainties,
alternative calibrations, or dataset substitutions. A common approach is to
duplicate parts of a workflow or to embed variation-specific logic directly
within the analysis itself. While effective, this couples the description of
the nominal analysis with the mechanics of evaluating systematic scenarios.

Flow instead treats variations as an independent layer on top of the workflow.
The nominal analysis describes the scientific computation once, while
variations specify only the modifications required to produce alternative
results. Variations may, for example, replace an input dataset, substitute an
object collection, or modify event weights, and can be restricted to specific
datasets or event types.

\begin{figure}[!htbp]
  \centering
  \includegraphics[width=\textwidth]{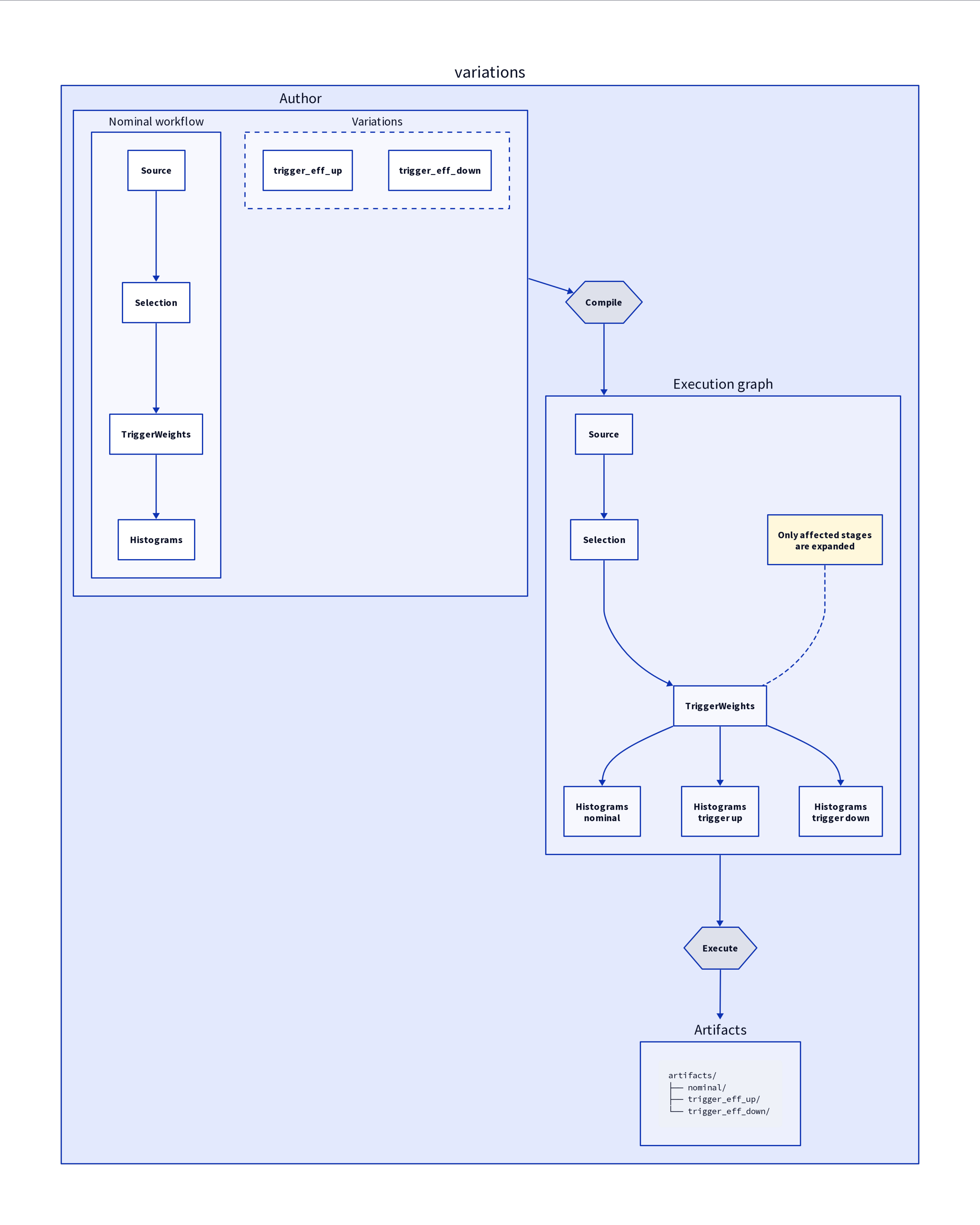}
  \caption{
    Variations are described independently of the nominal workflow and combined
    with it during compilation. Only stages affected by a variation are expanded
    into separate execution paths, while outputs are organised by variation.
  }
  \label{fig:variations}
\end{figure}

During compilation, the nominal workflow is combined with the selected
variations to produce the execution graph. Rather than replicating the complete
workflow, the compiler propagates each variation through the dependency graph
and expands only those stages whose inputs depend on the modified quantity, as
illustrated in Figure~\ref{fig:variations}. This explicit representation makes
the relationship between variations and the affected computation visible while
also providing a natural foundation for future execution optimisations.

Separating variations from the nominal workflow provides a convenient
model for analysis development. New systematic scenarios can be
added, modified, grouped or disabled without restructuring the
workflow itself. The explicit dependency information retained by the
compiler also enables future execution optimisations. For example, a
runtime may determine that the inputs to an unaffected stage are
unchanged and reuse a previously computed result rather than
executing the stage again.

Although execution may share computation internally, outputs are organised by
variation. This preserves a clear correspondence between each systematic
scenario and its resulting artifacts while maintaining complete provenance for
every execution path.

\subsection{Outputs and observation}
\label{sec:workflow-language-observation}

Flow separates scientific computation from execution observation and output
generation. Rather than embedding debugging, profiling or output generation
directly within operations, these concerns are described declaratively
alongside the workflow. This allows them to be added, modified or removed
independently of the analysis itself.

Observers inspect the execution of selected workflow stages without modifying
their behaviour. They provide a convenient mechanism for debugging, profiling
and validation by recording information about the data flowing through the
workflow. For example, the following observer records the schema of the data
after a particular analysis stage:

\begin{lstlisting}[style=yaml]
observers:
  - kind: hep.schema_snapshot
    at:
      - stage.SelectDimuonEvents
    out: schema
\end{lstlisting}

The observer produces a structured description of the data at runtime,
including the available fields, data types and partition metadata, allowing
analyses to be inspected without modifying the workflow itself.

While observers inspect execution, sinks consume the outputs of workflow
stages to produce user-visible artifacts such as ROOT files, tables, rendered
plots or reports. Sinks may execute immediately after a stage or be deferred
until the end of the workflow. The following example requests that a histogram
be rendered once the workflow has completed:

\begin{lstlisting}[style=yaml]
render:
  style: dimuon_mass
  when: final
\end{lstlisting}

Attaching outputs declaratively to workflow stages rather than writing files
directly from operations provides a consistent model for artifact generation
across different execution backends.

An additional consequence of this design is that provenance can be captured
automatically. Because each produced artifact is associated with a workflow
node, Flow records the information required to trace every output back to the
analysis that created it. This includes the producing workflow node, the input
dataset and partition, the compiled workflow and execution graph, together
with the software versions and execution environment used during the run.
Provenance therefore becomes an intrinsic property of the workflow rather than
an additional responsibility of the analysis author. Artifacts remain
reproducible and traceable without requiring additional instrumentation by the
analysis itself.

\subsection{From workflow to execution graph}
\label{sec:workflow-language-graph}

The workflow description introduced in this chapter is intended for authors
rather than execution. Before execution, Flow compiles it into progressively
more explicit intermediate representations, beginning with a normalised
workflow and ultimately producing a backend-independent execution plan.
Chapter~\ref{sec:compilation} describes this compilation process in detail.

\section{Compilation}
\label{sec:compilation}

The workflow language described in
Section~\ref{sec:workflow-language} presents an author-oriented view of an
analysis. Before execution, Flow transforms this description into a
backend-independent execution plan through a sequence of explicit compiler
passes. These passes resolve authoring conveniences, construct the logical
data-flow graph, analyse the data required by each operation, inspect the input
datasets, and divide the resulting computation into executable partitions.

The logical graph is introduced early in this process and remains the central
representation throughout compilation. Subsequent passes analyse and enrich
the graph rather than independently reconstructing the workflow. This makes
dependencies between operations, data products, observers and sinks explicit,
while providing a common representation on which applicability conditions,
variations and execution policies can operate.

Compilation produces inspectable artifacts at each stage. These include the
normalised workflow description, the logical graph, dependency reports,
discovered dataset metadata and the final execution plan. Intermediate
representations are therefore treated as first-class outputs rather than
transient implementation details. They allow compilation to be inspected,
validated and debugged independently of execution.

\subsection{Overview}
\label{sec:compilation-overview}

\begin{figure}[!htbp]
  \centering
  \includegraphics[width=\textwidth]{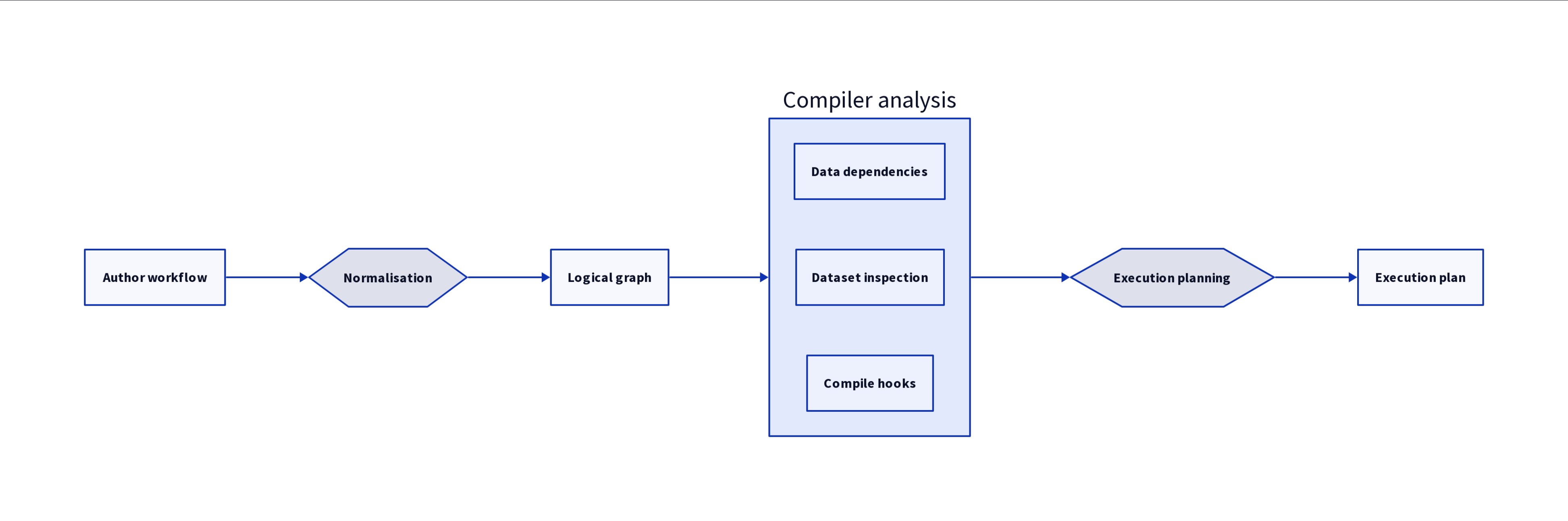}
  \caption{
    Flow compilation transforms the author workflow into a backend-independent
    execution plan. The workflow is normalised and lowered into a logical graph,
    which is subsequently enriched by dependency analysis, dataset inspection
    and other compiler passes before execution partitions are constructed.
  }
  \label{fig:compilation-pipeline}
\end{figure}

Figure~\ref{fig:compilation-pipeline} summarises the compilation process. The
author workflow is first normalised into a canonical representation. It is
then lowered into an explicit logical graph whose nodes represent sources,
transformations, observers and sinks, and whose edges identify the products
passed between them. Compiler analyses determine the fields required and
produced by each node, while dataset inspection obtains the structural
metadata needed to partition the input data. Finally, the logical graph and
the information gathered during compilation are combined into an execution
plan suitable for a runtime backend.

The distinction between the logical graph and the execution plan is important.
The graph describes the computation and the products exchanged between its
nodes, but does not prescribe how many times each node will execute. The plan
combines this graph with dataset metadata, partitioning policy and lifecycle
information to describe the concrete units of work presented to the runtime.

For the \(Z\rightarrow\mu\mu\) example, the logical graph is written to
\codepath{compile/analysis.graph.yaml}. The graph contains one source
node, three
analysis transformations and a rendering sink. Relationships that were
implicit in the author workflow are represented by explicit directed edges,
including the histogram passed from the final analysis stage to the rendering
operation. The execution plan subsequently expands this graph over the nine
input datasets and their corresponding partitions.

Active profiles determine the operations and compiler extensions available
during this process. Graph nodes retain stable registered implementation
identifiers, while compiler validation and planning ensure that the selected
sources, transformations, observers and sinks are compatible with the
execution environment. This allows the author workflow to remain independent
of concrete Python classes and backend-specific scheduling mechanisms.

The following sections describe the principal phases of compilation:
normalisation, lowering to the logical graph, dependency and dataset analysis,
execution planning, and the production of auxiliary compiler artifacts.

\subsection{Workflow normalization}
\label{sec:compilation:normalization}

Compilation begins by normalizing the author workflow. The purpose of
normalization is to assemble the workflow from all of its inputs and
expand abbreviated author syntax into a single, explicit
configuration. The resulting document is written to
\codepath{compile/normalised.yaml} and forms the input to all
subsequent compilation stages.

Workflow definitions need not reside in a single file. Authors may
separate reusable or independently maintained components into
additional files and include them from the primary workflow, for example:

\begin{lstlisting}[style=yaml]
include:
  - datasets.yaml
  - outputs.yaml
  - styles.yaml
\end{lstlisting}

These files may contribute datasets, output definitions, styles, or
other workflow sections. Selected profiles constitute another class
of input. Profiles may provide operation registries, execution hooks,
compiler extensions, renderers, report templates, and execution
configuration. During normalization, these contributions are merged
with the primary workflow to produce a complete configuration.

Normalization also expands compact author syntax into its canonical
representation. Dataset defaults are materialized for every dataset,
optional workflow sections are inserted with neutral values, and
profile contributions are expanded into their concrete definitions.
The normalised workflow therefore becomes entirely self-contained:
subsequent compiler phases operate on a single assembled document
without repeatedly loading includes, interpreting shorthand syntax,
or consulting profile definitions.

The normalised workflow additionally records the provenance of the
assembled configuration. Each contributing layer---including built-in
definitions, profiles, included files, and the primary workflow---is
recorded together with the registry entries it introduces. This
provenance allows users to determine where individual configuration
elements originate and provides a transparent view of the workflow
assembly process.

Normalization deliberately does not change the computational
structure of the workflow. Analysis stages remain in their
declarative author representation, and style inheritance expressed
through \codepath{use} and \codepath{with} is preserved. Likewise, graph
nodes and edges have not yet been constructed. These structural
transformations are performed during the lowering phase described in
the following section.

\subsection{Lowering to the logical graph}
\label{sec:compilation:lowering}

The normalised workflow retains the author-oriented structure of the analysis.
Stages, joins, observers, render declarations, and other workflow constructs
remain expressed using the declarative syntax provided by the author. Although
this representation is convenient for authoring, it is not well suited to
compiler analyses such as dependency resolution, validation, and execution
planning.

The next compilation phase lowers the normalised workflow to the \emph{logical
graph}, written to \codepath{compile/analysis.graph.yaml}. This graph
becomes the
central compiler representation for the remainder of compilation. Rather than
encoding the analysis as a sequence of workflow declarations, the logical graph
represents each computational component explicitly as a node connected by typed
data-flow relationships.

\begin{figure}[!htbp]
  \centering
  \includegraphics[
    width=\textwidth,
    height=0.82\textheight,
    keepaspectratio
  ]{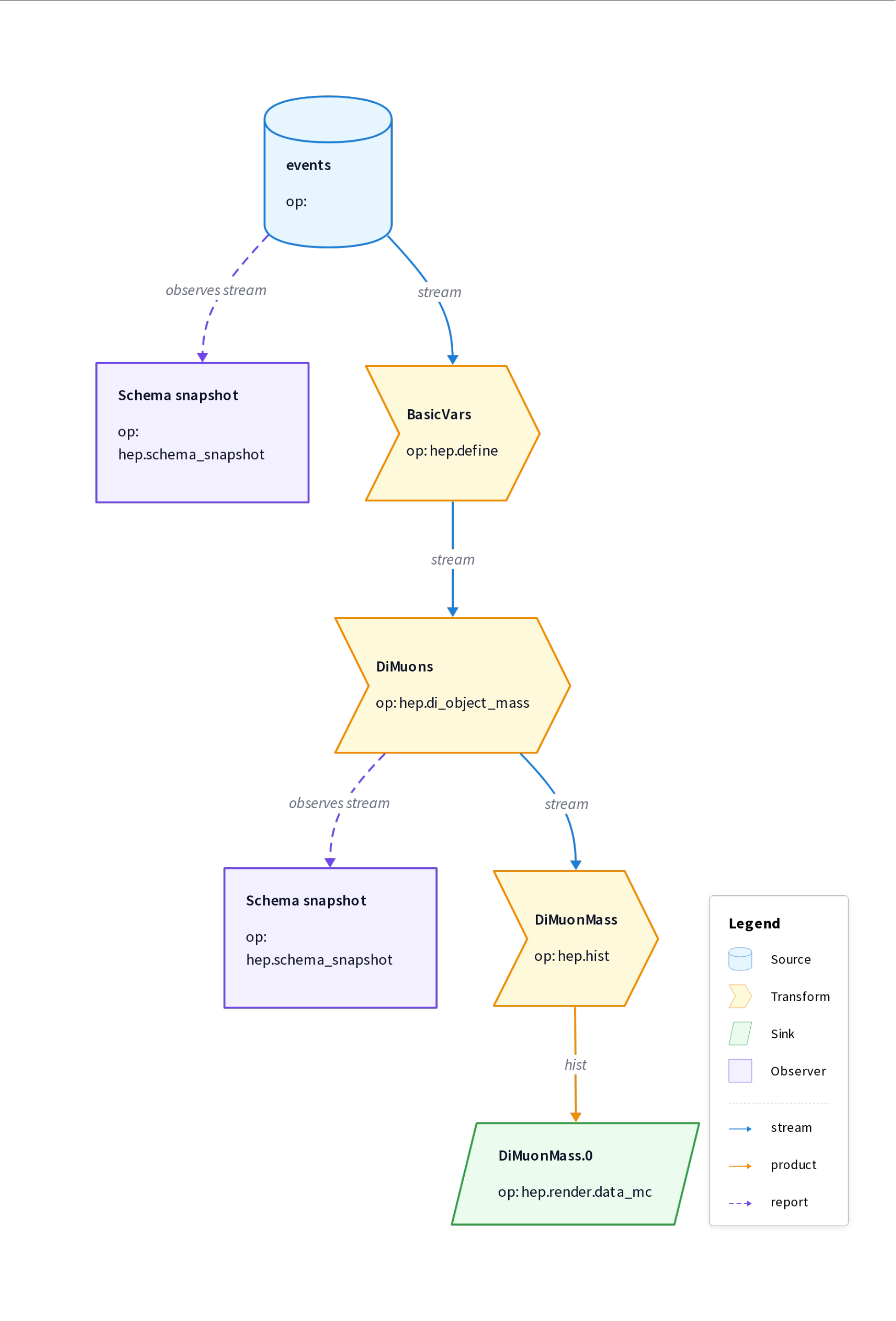}
  \caption{
    Lowered logical graph for the $Z \rightarrow \mu\mu$ analysis.
    Sources, transforms, sinks, and observers are represented as typed nodes;
    the operation implemented by each node is identified by the abbreviated
    label \codepath{op}.
    Edges distinguish event streams, derived products, and observer
    relationships.
    Schema observers attached after the source and the \codepath{DiMuons}
    transform expose the fields available at these graph boundaries and are
    discussed further in
    Section~\ref{sec:compilation:dependency-analysis}.
  }
  \label{fig:logical-graph}
\end{figure}

Figure~\ref{fig:logical-graph} shows the lowered logical graph generated for the
$Z \rightarrow \mu\mu$ analysis. Four principal node types are used throughout
the compiler:

\begin{itemize}
  \item \textbf{Sources} represent external data entering the workflow.
  \item \textbf{Transforms} perform computations on one or more input streams.
  \item \textbf{Sinks} consume streams or derived products to produce external
    outputs, such as files or rendered figures.
  \item \textbf{Observers} attach non-transforming analyses to streams,
    artifacts, or products. They leave the observed value unchanged,
    but may emit
    reports, metadata, or auxiliary products used for introspection, validation,
    provenance, or subsequent compilation stages.
\end{itemize}

The nodes are connected by typed edges representing different relationships.
Stream edges carry event data between computational stages, product edges carry
derived products such as histograms, and report or artifact edges represent
auxiliary information produced during compilation or execution.

Lowering converts author-level workflow constructs into these graph elements.
Workflow sources become source nodes, analysis stages become transform nodes,
render declarations become sink nodes, and observers are attached to the
appropriate graph boundaries. Sequential stages are converted into explicit
stream dependencies, while joins and other multi-input operations introduce
the corresponding graph connectivity.

Lowering also resolves several author-level abstractions. Render styles are
expanded into concrete renderer configurations, render variations are converted
into independent sink nodes, field aliases are resolved, and projection nodes
are inserted where downstream computations require only a subset of the
available fields. The resulting graph makes these relationships explicit while
preserving the semantics of the normalised workflow.

\begin{figure}[!htbp]
  \centering
  \includegraphics[
    width=\textwidth,
    height=0.72\textheight,
    keepaspectratio
  ]{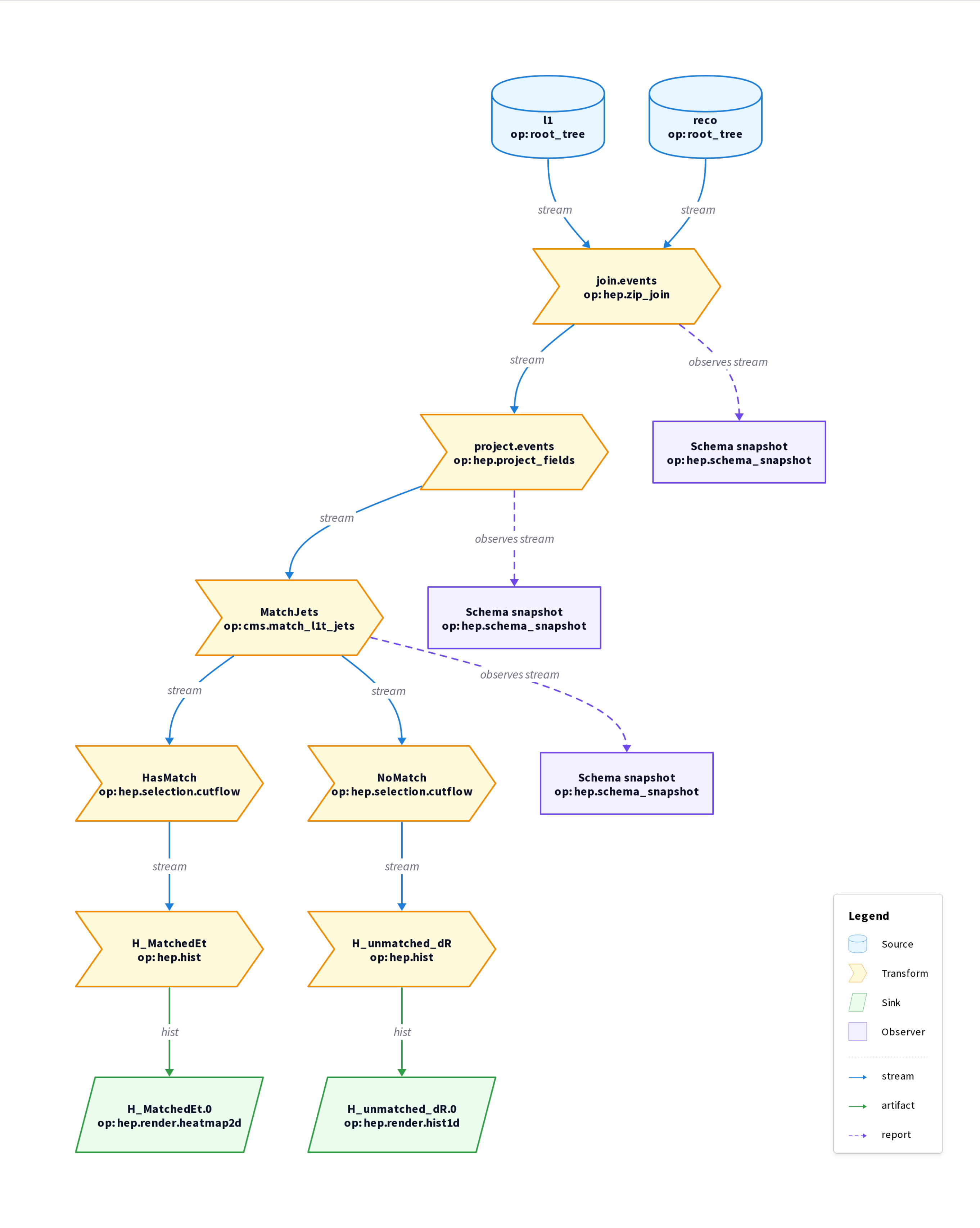}
  \caption{
    Multi-source logical graph for an analysis combining Level-1 and
    reconstructed-event data.
    The \codepath{join.events} transform combines two independent
    source streams,
    after which projection, matching, selections, histogramming, rendering, and
    schema observation proceed through the ordinary graph model.
    Explicit source and join nodes keep the workflow representation independent
    of any assumption that all event data reside in a single tree.
  }
  \label{fig:multi-source-logical-graph}
\end{figure}

The logical graph is not restricted to workflows with a single source.
Figure~\ref{fig:multi-source-logical-graph} shows an analysis that reads
independent Level-1 and reconstructed-event trees and combines them through an
explicit join node. Downstream projections, matching operations, selections,
histograms, renderers, and observers operate on the joined stream using the
same graph model as in a single-source workflow.

Supporting multiple sources and explicit joins is a deliberate property of the
intermediate representation. Data required by an analysis may be distributed
across separate files, trees, tables, or services, and these storage layouts may
evolve independently of the analysis itself. Restricting the workflow model to
a single input tree would embed a particular storage layout into the analysis
abstraction and make later evolution unnecessarily difficult. Flow instead
represents each source and the relationships between sources explicitly in the
logical graph.

The logical graph forms the common representation used by the remaining
compiler phases. Dependency analysis reasons about field requirements and
availability using the graph topology, validation verifies structural
correctness, execution planning partitions the graph into executable units, and
observers record metadata and other auxiliary products at selected graph
boundaries.

Observers are more than debugging facilities. In
Figure~\ref{fig:logical-graph}, schema observers are attached
after the source and after the derived di-muon stage. These record the schema
available at those points in the workflow and are used in the following section
to illustrate dependency propagation through the graph. Other observers may
produce machine-readable metadata rather than human-readable reports. For
example, the dataset inspection observer provided by
\codepath{fasthep-curator} generates the
\codepath{dataset_entries.json} artifact used during execution planning.

Observers and execution modifiers share a similar attachment model, but have
different semantics. Both may be associated with particular graph nodes or
graph boundaries. Observers are non-transforming: they inspect a stream,
artifact, or product without altering the value being observed. Execution
modifiers, in contrast, participate in runtime execution and may alter the data
flowing through a node, the behaviour of the node itself, or the environment in
which it executes. Maintaining this distinction separates passive analysis and
metadata collection from runtime intervention.

Once sources, joins, projections, and downstream computations have been made
explicit, the compiler can determine field dependencies independently for each
input stream. The following section describes this analysis.

\subsection{Dependency and dataset analysis}
\label{sec:compilation:dependency-analysis}

Following lowering, the compiler possesses a complete logical graph describing
the analysis topology, but not yet the precise symbols required to execute it.
The purpose of dependency analysis is to determine which symbols must be
available at every point in the graph, validate that these requirements can be
satisfied, and derive the minimal set of symbols that must be read from each
source. Unlike many analysis frameworks, this reasoning is performed entirely
during compilation before any event processing begins.

\begin{figure}[!htbp]
  \centering
  \includegraphics[
    width=\textwidth,
    height=0.32\textheight,
    keepaspectratio
  ]{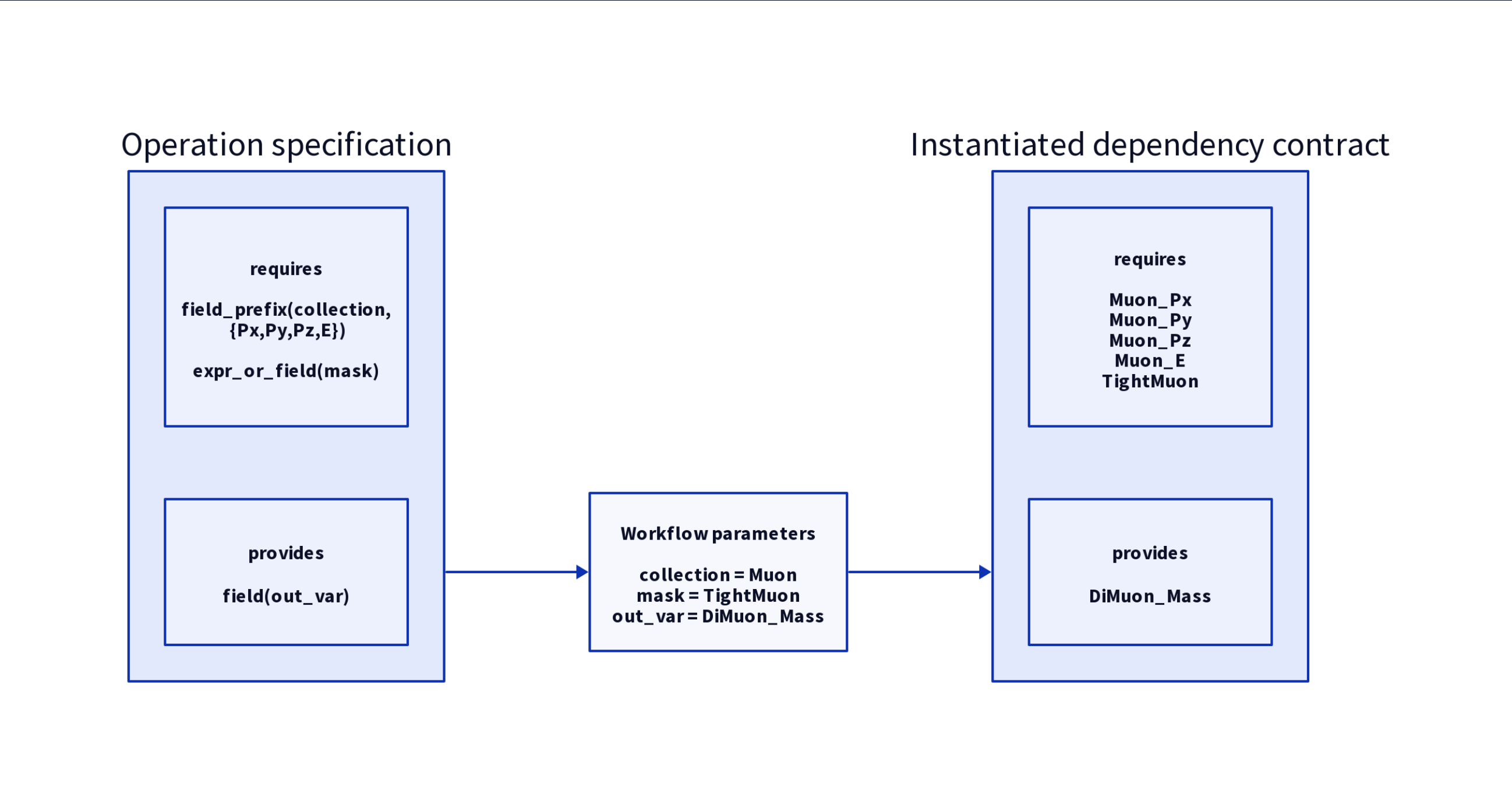}
  \caption{
    Instantiation of an operation dependency contract.
    Operation specifications declare symbolic input (\codepath{requires}) and
    output (\codepath{provides}) dependencies independently of any particular
    analysis.
    During compilation these symbolic dependencies are instantiated using the
    workflow parameters to derive the concrete symbols propagated through the
    logical graph.
  }
  \label{fig:dependency-contract}
\end{figure}

Dependency analysis is driven by the operation specifications rather than by
operation-specific compiler logic. Each operation declares a dependency
contract consisting of the symbols it requires from its inputs and the symbols
it provides to downstream stages. These declarations describe the intrinsic
data-flow semantics of the operation independently of any particular workflow.
During compilation, the contracts are instantiated using the parameters
supplied by the workflow to produce concrete symbol dependencies.
Figure~\ref{fig:dependency-contract} illustrates this process. Symbolic
dependencies declared by the operation specification are combined with the
workflow parameters to derive the concrete symbols used during dependency
propagation and validation.

As an illustrative example, the operation responsible for computing an
invariant mass does not declare dependencies on the \codepath{Muon} collection
explicitly. Instead, its specification requires the four-momentum components of
an arbitrary collection together with an optional mask expression, while
providing a configurable output symbol. When instantiated with the workflow
parameters
\codepath{collection = Muon},
\codepath{mask = TightMuon}, and
\codepath{out_var = DiMuon\_Mass},
the compiler derives the concrete dependencies
\codepath{Muon_Px},
\codepath{Muon_Py},
\codepath{Muon_Pz},
\codepath{Muon_E},
and
\codepath{TightMuon},
while determining that the operation produces the symbol
\codepath{DiMuon\_Mass}.
The compiler therefore remains entirely independent of physics-specific object
types, relying solely on the declared dependency contract.

Dependency propagation proceeds backwards through the logical graph. Sinks,
observers, and downstream transforms first declare the symbols they require.
Each transform then maps these requirements onto the symbols required by its
inputs according to its instantiated dependency contract. Symbols produced by a
node augment the schema available on its outgoing stream, while projection
operations explicitly reduce the available symbol set where appropriate. This
backward propagation continues until the minimal symbol requirements for every
source have been determined.

The dependency contracts also support symbolic expressions. Operations may
declare that a parameter represents either a field name or an expression over
fields. During compilation, such expressions are inspected statically to
identify every referenced symbol before dependency propagation proceeds. For
example, a selection such as
\codepath{Muon_Pt > 30 \&\& abs(Muon_Eta) < 2.4}
introduces dependencies on
\codepath{Muon_Pt} and
\codepath{Muon_Eta}
without requiring the operation itself to interpret the expression semantics.

The logical graph naturally extends this analysis to workflows containing
multiple independent sources. Each input edge of a multi-input transform
maintains its own dependency set, allowing requirements to be propagated
independently through each branch of the graph. In
Figure~\ref{fig:multi-source-logical-graph}, for example, the
\codepath{join.events} transform receives independent Level-1 and reconstructed
event streams. The compiler therefore derives separate source projections for
the two inputs while treating the join as an ordinary graph transform rather
than a special compilation case.

One consequence of performing dependency analysis during compilation is that
schema validation can also be completed before execution. Once the required
symbols for each source have been determined, they are compared with the schema
available from the corresponding dataset. Missing fields, incompatible
dependencies, or unresolved symbols therefore produce compilation errors rather
than runtime failures after computational resources have already been consumed.
The same dependency information is subsequently used to configure source
projections so that only the symbols required by downstream computation are
materialized during lazy data loading.

Dependency analysis concerns the logical contents of the event streams, while
execution planning additionally requires information about the physical
datasets themselves. Dataset inspection is therefore represented as an observer
attached to the relevant source nodes. Unlike schema observers, which expose
the evolution of symbols throughout the graph, dataset observers collect
metadata about the underlying datasets, including entry counts, available
schemas, and other source properties required during planning. In the current
implementation this functionality is provided by
\codepath{fasthep-curator}, which generates the
\codepath{compile/dataset_entries.json} artifact consumed by the execution
planner.

The outputs of this phase include the required symbols for every source, the
symbols available at each graph boundary, validation diagnostics, source
projections inserted by the compiler, and dataset metadata required for
execution planning. Together these analyses transform the logical graph from a
structural representation of the workflow into a validated, data-aware
representation from which execution plans can be derived.

\subsection{Compilation extensions and artifacts}
\label{sec:compilation:extensions}

The compilation pipeline described in the preceding sections defines the
standard transformation from a declarative workflow to an executable plan.
Flow does not, however, require every compilation concern to be implemented
within the core compiler.
Compilation may be extended by external packages that inspect, enrich, or
transform intermediate compiler representations and emit additional
machine- or human-readable artifacts.

This mechanism is already used by several components of the FAST-HEP
ecosystem.
Dataset inspection is provided by
\codepath{fasthep-curator}, which examines the physical dataset inputs and
collects metadata such as the files associated with each dataset and the
number of entries available in each file.
These results are recorded in
\codepath{compile/dataset_entries.json}
and are subsequently used during dependency analysis and execution planning,
for example when constructing file- and entry-range partitions.

Graph rendering is provided separately by
\codepath{fasthep-render}.
It consumes the logical graph produced during compilation and emits a visual
representation such as
\codepath{graph/graph.svg}.
The rendered graph is not required for execution, but provides a useful
inspection and debugging artifact for users and developer tools.

These examples illustrate two complementary forms of compilation extension.
An extension may contribute information required by later compiler phases, as
in dataset inspection, or it may consume compiler state to produce an
auxiliary artifact, as in graph rendering.
In both cases, the extension operates through explicit compiler interfaces
rather than by embedding package-specific behaviour in the core compilation
pipeline.

The principal compiler artifacts currently include:

\begin{description}
  \item[\codepath{compile/normalised.yaml}]
    the assembled and normalised workflow definition;

  \item[\codepath{compile/analysis.graph.yaml}]
    the lowered logical graph;

  \item[\codepath{compile/dataset_entries.json}]
    metadata collected from the configured physical datasets;

  \item[\codepath{compile/plan.yaml}]
    the backend-independent execution plan; and

  \item[\codepath{graph/graph.svg}]
    a rendered representation of the logical graph.
\end{description}

Only the execution plan is required by the runtime.
The remaining artifacts expose intermediate compiler state for inspection,
validation, visualization, debugging, provenance, and integration with
external tooling.

The extension mechanism is intended to support deeper customization as it
matures.
Future extensions may participate earlier in compilation by introducing
additional normalization passes, resolving new forms of workflow abstraction,
or translating alternative authoring syntaxes into Flow's normalised workflow
representation.
This would allow domain-specific or experiment-specific workflow languages to
reuse the logical graph, dependency analysis, execution planning, and runtime
infrastructure without requiring the core compiler to understand their
surface syntax.

Consequently, the normalised workflow and logical graph serve not only as
internal compiler representations, but also as extension boundaries.
An external frontend may produce the normalised representation expected by
the compiler, while additional compiler passes may inspect or transform that
representation before lowering and planning continue.
The standard Flow syntax is therefore one frontend to the compilation
pipeline rather than an intrinsic requirement of the runtime architecture.

The long-term objective is a compiler architecture in which functionality can
be distributed across focused packages---for example, dataset inspection in
Curator and visualization in Render---while preserving a stable compilation
model and a common execution-plan interface.

\section{Execution}
\label{sec:execution}

The previous chapter described how Flow compiles a declarative workflow into
the backend-independent execution plan
\codepath{compile/plan.yaml}.
This chapter describes how the runtime realises that plan on local or
distributed computing resources.

The execution plan forms the contract between compilation and execution.
Compilation resolves workflow semantics, dependencies, partitioning, lifecycle
assignment, and operation configuration. The runtime interprets the resulting
plan, while the selected backend determines where the planned work is carried
out.

\begin{figure}[!htbp]
  \centering
  \includegraphics[width=\textwidth]{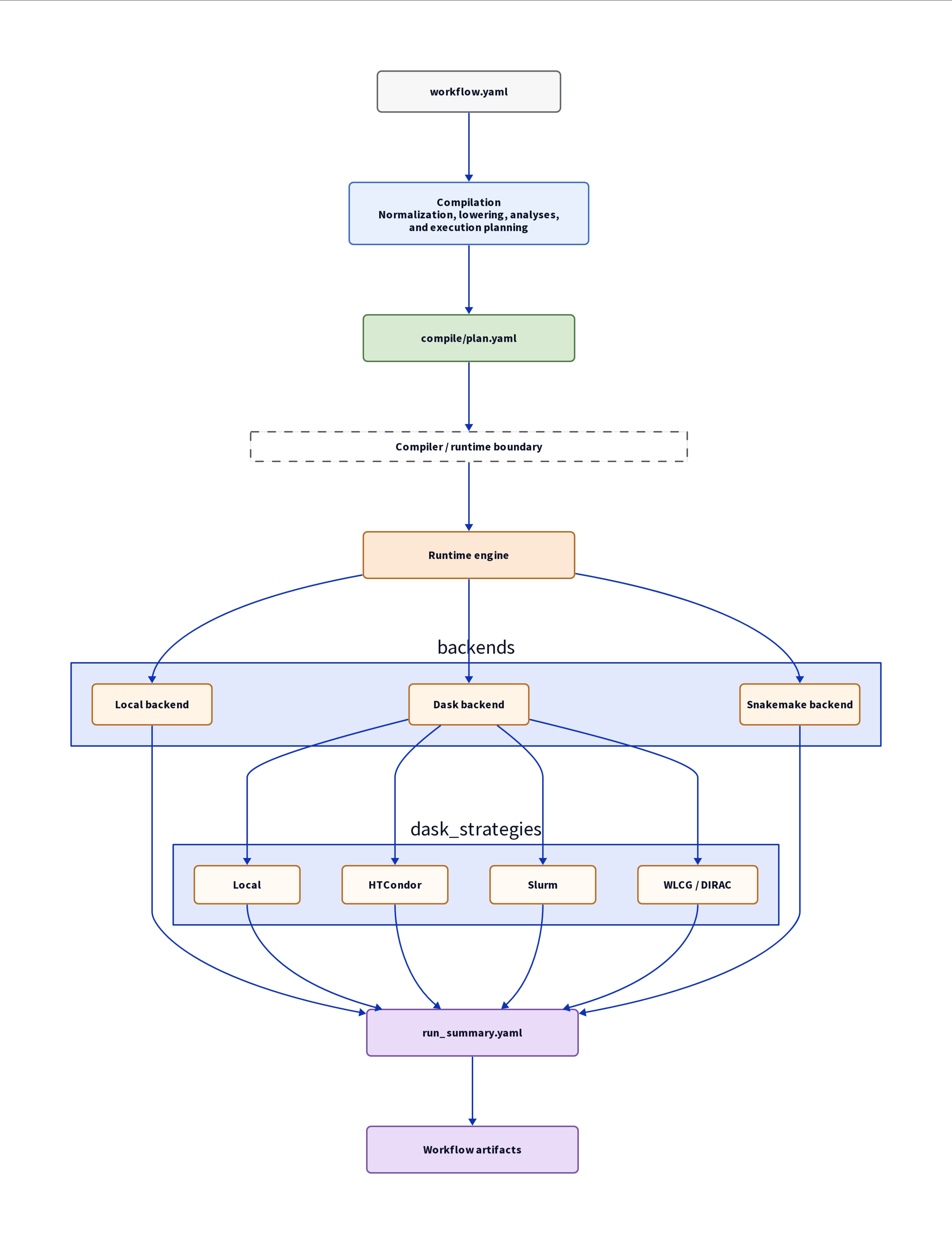}
  \caption{
    Overview of the execution architecture.
    Compilation produces the backend-independent execution plan
    (\codepath{compile/plan.yaml}),
    which is interpreted by the runtime engine and executed through the
    selected backend.
    Execution produces a runtime summary
    (\codepath{run_summary.yaml})
    together with the workflow artifacts.
  }
  \label{fig:execution-pipeline}
\end{figure}

Figure~\ref{fig:execution-pipeline} summarises this architecture. A common
runtime engine instantiates the execution plan and coordinates lifecycle
transitions, product handling, and provenance recording. Backend
implementations supply the computing resources and scheduling mechanism.

\subsection{Runtime engine}
\label{sec:execution:runtime}

The runtime engine is responsible for instantiating the execution plan.
Unlike the compiler, it performs no analysis or optimization of the workflow.
Instead, it interprets the compiled execution plan and realizes the execution
described therein.

For each execution partition, the runtime constructs the appropriate execution
context, resolves the planned implementation identifiers through the
configured runtime
registries, binds the required inputs, evaluates the planned execution nodes,
dispatches lifecycle hooks, records provenance information, and materialises
the resulting products. Products remain associated with the execution
nodes and partitions that
produced them and are subsequently merged according to the lifecycle encoded
during compilation.

Because all analysis-dependent decisions have already been resolved during
compilation, the runtime engine remains independent of individual workflows
and operation semantics. New operations are introduced through runtime
registries rather than modifications to the execution engine itself.

\subsection{Execution lifecycle}
\label{sec:execution:lifecycle}

Execution proceeds hierarchically across three lifecycle scopes:
partition, dataset, and workflow.

Partition-level operations execute independently for each execution partition,
allowing data processing to be parallelised across computing resources.
Products generated during partition execution are then merged within each
dataset before dataset-level operations execute. Workflow-level operations
execute once after all datasets have completed.

This hierarchy allows reductions, report generation, rendering, and publication
to occur only after their required inputs are available. Lifecycle assignment
is determined during compilation and encoded in the execution plan, so the
runtime applies the prescribed lifecycle without re-analysing the workflow.

\subsection{Execution backends}
\label{sec:execution:backends}

Execution backends provide the computational resources required to execute the
compiled workflow. Their responsibility is intentionally limited to scheduling
execution partitions, allocating resources, and collecting execution results.
Workflow semantics, operation execution, lifecycle management, provenance, and
product handling remain entirely within the shared runtime engine.

This separation allows the same execution plan to execute unchanged across
multiple computing environments. Operational parameters such as
worker counts, queue selection, and resource
limits may be supplied when execution begins without changing the semantic
contents of the execution plan. Any such overrides are recorded in the runtime
summary.

The current implementation provides a sequential local backend intended for
interactive development together with a distributed backend based on Dask.
The local backend executes execution partitions sequentially using the shared
runtime engine. The Dask backend submits the same partition executions as
distributed tasks while reusing the identical runtime implementation for
partition execution, lifecycle transitions, provenance recording, and product
merging. The distributed backend is based on Dask. It may connect directly to an
existing Dask scheduler or provision workers through batch systems such as
HTCondor and Slurm. These provisioning strategies affect resource acquisition
but not runtime semantics.

\subsection{Execution artifacts}

Execution produces the runtime summary,
\codepath{run_summary.yaml},
which complements the execution plan generated during compilation.

Whereas \codepath{compile/plan.yaml} records the execution intended by the
compiler, \codepath{run_summary.yaml} records how that plan was
realised. It captures the selected execution backend and
strategy, processed partitions, generated products, warnings, hook activity,
execution statistics, provenance information, and references to the generated
workflow artifacts.

Together, the execution plan and runtime summary provide a complete record of
both the intended and realized execution of the workflow, supporting
reproducibility, validation, and debugging.

\subsection{Provenance}
\label{sec:execution:provenance}

Workflow execution produces not only analysis outputs, but also structured
provenance describing how those outputs were obtained. Provenance connects each
artifact to the compiled workflow, the execution node and partition that
produced it, the input data, and the software and runtime environment used
during production.

In the current implementation, provenance is recorded alongside the workflow
artifacts because its primary purpose is to describe their production. The
provenance output consists of three components:

\begin{description}
  \item[\codepath{artifacts/provenance/execution.json}]
    describes the workflow execution as a whole;

  \item[\codepath{artifacts/provenance/manifest.json}]
    indexes the generated artifacts and their provenance records; and

  \item[\codepath{artifacts/provenance/records/}]
    contains the individual provenance record associated with each artifact.
\end{description}

The execution record assigns a unique identifier to the run and identifies the
compiled workflow representations, software versions, runtime environment,
execution partitions, and workflow-node executions involved. It therefore
establishes a connection between the declarative workflow, its compiled form,
and the concrete units of work realised by the runtime.

The provenance manifest provides the entry point for artifact-level inspection.
For each recorded artifact, it identifies the artifact path and kind, the
workflow node that produced it, and the location of its individual provenance
record. A cryptographic hash identifies the record independently of its
filename and provides a basis for detecting accidental modification.

Each artifact record describes the artifact together with its immediate
producer and inputs. The producer is identified by the workflow node,
execution partition, and node-execution identifier. Input lineage is currently
recorded primarily at partition granularity. Together, the execution record,
manifest, and artifact records connect the compiled workflow to the concrete
executions and outputs realised by the runtime.

The current implementation records provenance for local execution. In this
case, the hostname, operating-system platform, Python interpreter, working
directory, and installed software versions provide a useful description of the
runtime environment. Distributed execution requires finer-grained provenance
because different partitions may be processed on heterogeneous workers with
different processor architectures, operating-system images, software
environments, accelerators, or numerical libraries. Recording only the
submitting host would therefore be insufficient.

Future distributed provenance must associate runtime conditions with
individual partition or node executions. This distinction matters for
scientific reproducibility as well as operational debugging: otherwise
identical workloads may produce small differences when executed using
different software builds, numerical libraries, or processor architectures.
Without worker-level provenance, diagnosing such differences after production
can be prohibitively difficult.

The same model can be extended beyond the immediate execution environment.
Useful scientific provenance includes the precise dataset versions, files,
objects, and fields that were read; the corrections, calibrations, and scale
factors that were applied; and the parameters and software implementations
used by each operation. These relationships would provide a trace from an
analysis artifact through the transformations and data products that
contributed to it.

The long-term objective is to generate human- and machine-readable provenance
reports capable of answering questions such as which inputs contributed to an
artifact, which corrections were applied, which runtime conditions were
present, and which software and data versions were used. The same information
may also support validation, caching, reuse, performance analysis, resource
accounting, and environmental reporting.

Provenance is currently implemented within Flow while its data model and
runtime integration are established. Once these interfaces stabilise, the
shared provenance representation and reporting facilities are intended to move
to Curator. Flow will remain responsible for emitting execution events and
artifact relationships, while Curator will assemble, store, and present the
resulting provenance records.

\section{Experience and lessons learned}
\label{sec:lessons}

The architecture presented in this paper did not emerge from a single design
exercise. Rather, it reflects the cumulative experience gained during the
development of FAST and its evolution into the FAST-HEP ecosystem over nearly
a decade. Many of the design decisions described in the preceding chapters
were motivated by practical limitations encountered while supporting real
analyses, evolving software dependencies, and maintaining the framework across
multiple generations of contributors.

One of the earliest goals of FAST was to replace imperative analysis code with
a declarative workflow description. This substantially reduced the amount of
analysis-specific software that physicists were required to write and improved
the readability of many analyses. Experience showed, however, that replacing
code with configuration did not by itself reduce the overall complexity of an
analysis. As analyses matured, configuration gradually became distributed
across multiple files describing datasets, corrections, systematic
uncertainties, histogram definitions, execution settings, and visualization
options. While each component remained declarative, understanding exactly what
would execute required reconstructing the complete analysis from many separate
sources.

The introduction of an explicit compilation stage fundamentally changed this
model. Rather than executing directly from the author-facing descriptions,
Flow consolidates all workflow inputs into a normalised representation before
performing graph construction, dependency analysis, and execution planning.
The resulting execution plan provides a single, authoritative description of
the intended analysis, while execution summaries and provenance records
describe how that plan was ultimately realized. Compilation therefore serves
not only to prepare execution but also to make analyses explicit and
inspectable, reducing ambiguity during validation, debugging, and long-term
maintenance.

A second recurring lesson was that declarative workflows alone do not
guarantee modular software. The original FAST framework supported extensive
customization through Python, allowing users to replace components or extend
existing functionality. In practice, however, the absence of explicit
contracts meant that many assumptions remained embedded within the framework
itself. Missing products, incompatible implementations, or inconsistent
configuration frequently surfaced only during execution, often after
substantial processing had already been performed. Flow instead validates
dependencies during compilation using the capability contracts declared by
operation specifications. Errors that previously appeared late during runtime
are now detected before execution begins, providing earlier feedback and
making workflow behaviour substantially more predictable.

Experience also demonstrated that replaceability cannot be achieved solely
through careful implementation; it must be reflected in the software
architecture. Throughout the development of FAST-HEP, replacing an underlying
library often required coordinated changes across otherwise unrelated
components because the framework implicitly encoded assumptions about internal
representations. Histogram accumulation provides a representative example.
Earlier versions relied heavily on multi-index \texttt{pandas} dataframes as
the intermediate histogram representation. While initially flexible, this
approach proved increasingly inefficient for large analyses and effectively
coupled the framework to a particular implementation strategy. Replacing the
histogram library or experimenting with alternative implementations required
changes throughout the framework rather than a simple substitution.

Flow instead treats histogram accumulation and merging as behaviour provided
by the histogram implementation itself. The runtime invokes the merge contract
associated with the registered histogram capability without assuming any
particular internal representation. Consequently, a new histogram
implementation can be introduced simply by providing the required behaviour
and registering it with the framework. Neither the compiler, runtime, nor
analysis descriptions require modification. Similar architectural boundaries
now exist for data sources, output formats, execution backends, compiler
extensions, and other replaceable components. This substantially lowers the
cost of evaluating new technologies as they emerge.

The same architectural principles enabled functionality that would have been
difficult to accommodate within the earlier framework. Representing workflows
as explicit graphs with well-defined source, transform, sink, and observer
nodes allows analyses to combine multiple independent data sources without
introducing special-case execution logic. Likewise, treating output generation
as graph sinks rather than framework-specific writers permits additional
output formats to be introduced without modifying the execution engine.
Supporting ROOT ntuples, alternative columnar formats, or future storage
technologies therefore becomes a matter of providing new implementations
rather than extending the core framework. This flexibility was a long-standing
objective of the original FAST project and has become a practical consequence
of the current architecture.

Another important lesson concerned the relationship between scientific
concepts and workflow language design. During several redesign iterations
there was a natural temptation to expose experiment-specific concepts directly
within the workflow syntax. Experience showed that this approach inevitably
expanded the core language while making it increasingly difficult to support
new domains. Flow instead keeps the workflow language deliberately compact,
allowing operation specifications to expose domain-specific semantics through
parameters and capability contracts. The core language therefore remains
stable while individual scientific domains retain the freedom to define their
own operations and associated compilation behaviour.

Perhaps the most significant lesson concerns the role of explicit
representations. Concise author-facing workflow descriptions are valuable, but
they are insufficient for understanding precisely what will execute or how a
particular scientific result was obtained. Flow therefore makes successive
stages of the compilation and execution process visible through normalised
workflow descriptions, logical graphs, execution plans, runtime summaries, and
provenance records. Rather than competing goals, concise authoring and
explicit execution descriptions have proven to be complementary. The former
reduces the burden on analysts, while the latter provides the transparency
required for validation, reproducibility, debugging, and long-term
preservation.

Finally, experience developing FAST-HEP largely through volunteer
contributions and relatively short periods of dedicated development effort
highlighted that software architecture directly affects contributor
sustainability. In tightly coupled systems, seemingly local improvements often
require understanding and modifying unrelated parts of the framework,
discouraging experimentation and increasing the cost of maintenance. Several
promising ideas were ultimately abandoned because evaluating them required
framework-wide changes rather than isolated implementations. The current
architecture therefore treats inexpensive experimentation as a primary design
requirement. New algorithms, execution technologies, storage formats,
instrumentation, or compiler analyses should be introducible through well-
defined extension points without modifying production analyses or core
framework components.

This philosophy also reflects the realities of high-energy physics analyses,
which frequently outlive the involvement of their original authors. Preserving
an analysis therefore requires preserving not only source code but also the
operational knowledge surrounding datasets, corrections, implementations,
execution environments, and produced artifacts. By making these aspects
explicit through compilation and provenance, Flow aims to reduce reliance on
implicit knowledge held by individual developers and to improve the
maintainability, reproducibility, and longevity of scientific analyses.

\section{Conclusions}
\label{sec:conclusions}

This paper presented the design and architecture of Flow, a
declarative workflow language, compiler, and
runtime designed to support high-energy physics analyses while separating
scientific intent from implementation details. Rather than interpreting
author-facing workflow descriptions directly, Flow employs an explicit
compilation pipeline that normalizes workflows, constructs a logical graph,
performs dependency analysis, and produces backend-independent execution
plans. A common runtime then executes these plans while recording execution
summaries and provenance.

A central objective of the architecture is to make workflow behaviour
explicit. Intermediate compiler representations, execution plans, and
provenance records expose information that would otherwise remain
implicit within configuration files, software implementations, and
execution environments. This improves transparency during
development, facilitates
validation and debugging, and contributes to the long-term reproducibility and
maintainability of scientific analyses.

The architecture also emphasizes replaceability through well-defined
behavioural contracts rather than fixed implementations. Operation
specifications, capability registries, execution backends, compiler
extensions, and workflow graph components provide stable interfaces through
which new technologies can be evaluated without requiring changes to existing
analysis descriptions or the core framework. This reduces the cost of
experimentation and allows the surrounding software ecosystem to evolve
independently of individual analyses.

The development of Flow has been strongly influenced by nearly a decade of
experience with FAST and FAST-HEP. Many of the architectural decisions
described in this paper arose directly from practical challenges encountered
while supporting evolving analyses, software libraries, and contributor
communities. The resulting architecture reflects these experiences by treating
explicit representations, modular compilation, and extensibility as primary
design objectives rather than implementation details.

Although the current implementation already supports production analyses,
Flow remains an active area of development. Future work will focus on
extending compilation through additional frontend and analysis
extensions, expanding execution capabilities for heterogeneous and
distributed computing environments, and further developing the
provenance model. As these interfaces mature, functionality currently
implemented within Flow, such as provenance management, is expected
to migrate into complementary ecosystem projects including Curator.

Rather than viewing workflows simply as configuration interpreted at
runtime, Flow treats them as programs that can be compiled, analysed,
validated, and executed through explicit intermediate
representations. This perspective has shaped the architecture
presented in this paper and provides a foundation for workflow
systems that can evolve alongside scientific analyses while remaining
transparent, extensible, and reproducible.

\section*{Declaration of AI-assisted work}
\label{sec:ai-use}

Use of Artificial Intelligence. Artificial intelligence tools were
used during the development of both the software and this manuscript.
For the manuscript, AI-assisted tools were used to generate initial
draft text from existing documentation and source code, and to
proof-read and improve the clarity of edited text. All technical
content, architectural decisions, and final wording were reviewed and
approved by the authors.

During software development, AI tools were primarily used as a form
of rubber-duck programming to discuss design ideas, explore
alternative approaches, and refine architectural decisions. AI was
also used to assist with the generation of boilerplate code and
routine refactoring tasks. All resulting code was reviewed, tested,
and integrated by the authors.

\section*{Acknowledgements}

The modernisation of FAST-HEP during 2021--2022 was supported by an IRIS
Digital Research Infrastructure (DRI) Funding Award, ``IRIS Digital Assets --
FAST-HEP: Faster Analysis Software Tools for HEP'', funded by STFC and awarded
to the University of Bristol.

\bibliographystyle{unsrt}
\bibliography{references}

\end{document}